\documentclass[preprint,aps,a4paper, showpacs]{revtex4}
\usepackage{graphicx}

\begin{document}
\draft
\title{Stochastic Fluctuations in Epidemics on Networks}
\author{M. Sim\~oes, M. M. Telo da Gama and A. Nunes 
\footnote{corresponding author \\ anunes@ptmat.fc.ul.pt}}
\address{Centro de F{\'\i}sica Te{\'o}rica e Computacional and 
Departamento de F{\'\i}sica, Faculdade de Ci{\^e}ncias da Universidade de 
Lisboa, P-1649-003 Lisboa Codex, Portugal}

\begin{abstract}

The effects of demographic stochasticity  in the long term behaviour of endemic 
infectious diseases have been considered for long as a necessary addition 
to an underlying deterministic theory. The latter would explain the regular
behaviour of recurrent epidemics, and the former the superimposed noise
of observed incidence patterns. Recently, a stochastic
theory based on a mechanism of resonance with internal noise
has shifted the role of stochasticity closer to the center stage, 
by showing that the major dynamic patterns found in the 
incidence data can be explained as resonant fluctuations, 
whose behaviour is largely independent of the amplitude of
seasonal forcing, and by contrast very sensitive to the basic epidemiological
parameters. Here we elaborate on that approach, by adding an 
ingredient which is missing in standard epidemic models, the 
'mixing network' through which infection may propagate.
We find that spatial correlations have
a major effect in the enhancement of the amplitude and the coherence
of the resonant stochastic fluctuations,
providing the ordered patterns of recurrent epidemics,
whose period may differ significantly from that of the small
oscillations around the deterministic equilibrium. 
We also show that the inclusion of a more realistic, time
correlated, recovery profile
instead of exponentially distributed infectious periods  may, even
in the random-mixing limit, contribute to the same effect. 
\end{abstract}


\maketitle

\section{introduction}

The incidence patterns of childhood diseases in the twentieth century 
have been a challenge and a preferred testing ground for 
epidemiological models. During the last decade, more sophisticated
approaches building on the traditional SIR and SEIR models \cite{amay}
 (see Section II for a description) 
have brought considerable advances in understanding and selecting 
some of the fundamental ingredients of the complex dynamics of infectious
diseases (\cite{hethcote}, \cite{Grenfnat2001},  \cite{wang}). 
The interplay between the system's nonlinearity and the periodic 
perturbation in seasonally forced SIR and SEIR models was shown to
be a source of complex dynamics compatible with the diversity 
of observed incidence records (\cite{Grenfsci2000}, \cite{b&earn}, 
\cite{DushoffPNAS}).  

This body of work belongs to an essentially deterministic framework,
where demographic stochasticity plays a secondary role, except when 
addressing stochastic extinction (\cite{bartlett},\cite{ccsize}). In this framework, the
role of stochasticity is that of sustaining small amplitude oscillations
around the deterministic system's equilibrium that follow the 
natural frequency given by the local linear 
approximation (\cite{b&earn}, \cite{bailey}, \cite{lloyd2004}), 
or else that of promoting the switching between different competing attractors of the 
underlying deterministic model (\cite{GrenfpD2001},\cite{vddrische}).

Recently, a stochastic theory developed for a predator-prey competition
model \cite{mckane2} and then applied to the well-mixed SIR model \cite{mckane1} has shown 
that the fluctuation power spectrum of the incidence time series is
essentially determined, both in the presence and in the absence of
seasonal forcing, by the resonance with internal noise of the system's 
natural frequency in the deterministic (infinite population size) limit. 
The cross correlation structure of this internal
noise can be computed analytically from the transition probabilities,
and it may shift the resonance peak away from the system's natural frequency.
The amplitude and the coherence of these resonant fluctuations are
both large for the parameter values that correspond to some
childhood diseases, so that they may be comparable  even in large systems
to the oscillations induced by seasonal forcing.

Apart from stochasticity, another missing ingredient of the standard epidemic models 
that has been attracting increasing attention
is the host population contact structure, or 'mixing network' \cite{keelingrev},
\cite{mixingpatterns}.
Most of the research connecting networks and epidemiology dates from the last five years, 
when many fundamental results of network theory became widely known while new ones have
been derived \cite{p41}, \cite{p42}, \cite{p43}. These ideas originated in the mathematics and physics communities 
were applied in an epidemiological setting from the beginning \cite{w&s}, 
\cite{p52}, \cite{p53} prompting the 
interest of epidemiologists and several theoretical \cite{p61},
\cite{p62}, \cite{p63}  and field epidemiology 
\cite{p71}, \cite{p72}, \cite{p73} results. 

Here we address the effect of relaxing the random mixing assumption on the
behaviour of the resonant fluctuations of a stochastic SIR model. 
We assume that the host population contact network may be represented by
a 'small-world' network of the type introduced by Watts and Strogatz nearly a decade ago \cite{w&s}. 
Analysis of real networks of social contacts \cite{realnets} indicates that this 
is a reasonable
first assumption to model the contact structure relevant for the propagation
of airborne infections. We find that spatial correlations have
a major effect in the enhancement of the amplitude and of the coherence
of the resonant stochastic fluctuations, providing ordered patterns of recurrent epidemics. 
The shift of the resonance peak away from the system's natural frequency 
also increases significantly due to correlations. The enhanced amplitude
and coherence of the fluctuations implies that in populations 
where a disease spreads predominantly through local infectious contacts,
large epidemic outbursts with well defined recurrence times
are generated by internal noise alone, without the presence of seasonal 
forcing. This effect is illustrated in Section II.B with a particular
example, and studied systematically in Section III.B.   

Another assumption of the SIR model  which has 
been challenged  \cite{keelingccsize}, \cite{lloyd2001}, \cite{wearing2005} 
is that of considering a constant recovery probability during the infection period. We 
study the effect on the resonant fluctuations of switching from the
(uncorrelated) exponentially
distributed infection periods used in \cite{mckane1} 
to the opposite limit of (time correlated) constant infection
periods. Again, we find that the spectrum of stochastic fluctuations 
found in \cite{mckane1} changes under more realistic assumptions on the 
recovery profile in such a way that the oscillations become larger and
more sharply defined. In this case, the peak frequency shift with respect to
the deterministic prediction is larger, and towards larger fequencies.

Qualitatively, the effects of spatial and time correlations on the fluctuation
spectrum are similar: both the spatial correlations, introduced through the
host population contact network, and the time correlations, introduced via
constant infection periods, lead to enhanced more sharply defined dominant
peaks. There are analogues of this behaviour in many systems studied in
statistical physics.
The peak frequency shift, however, is more pronounced and has a different
sign for the model with constant infection period. 
As discussed in Sections II.B and III.C, this qualitative 
difference can be understood on the basis of an effective deterministic 
description.

These results show that the classical understanding of the incidence 
time patterns of endemic infectious diseases, which is mainly based on a
seasonally forced deterministic description, is clearly insufficient 
to have a correct description of such fluctuations. 
These may be purely stochastic, and the diversity of incidence patterns found in
real data for the same disease in different populations can be
understood in this framework as an effect of population size and
contact structure.

Moreover, the approximate period of the recurrent epidemics
driven by internal noise in the presence of spatial and/or time
correlations differs significantly from the period computed in the usual 
deterministic approach. This is an important cautionary note with 
regard to the use, in the framework of the deterministic
description, of the recurrence period to help assess estimates
of epidemiological parameters. The breakdown of the assumptions of random mixing 
of the population 
and/or of constant recovery rate during the infectious period  
implies important corrections to the dominant frequency of the fluctuations.

\section{methods}

\subsection{Susceptible-Infectious-Recovered (SIR) dynamics in a randomly mixed
discrete population}

One of the simplest epidemiological models one can consider
is based on dividing the whole population in three classes of individuals, 
the susceptible, the infectious, and the permanently recovered.
It is suitable to study the infection dynamics of diseases that confer
long lasting immunity, such as childhood infections, provided that we take into
account in the long term dynamics the replenishment of susceptibles in
the population through births. 

The deteministic description of the basic Susceptible-Infectious-Recovered (SIR)
model in a closed population where renewal of susceptibles occurs through births
with a constant birth rate and death rate $\mu $  is given by the equations
\begin{eqnarray}
\dot s & = & \mu (1- s)-  \beta s i \nonumber \\
\dot i & = & \beta s i - (\gamma + \mu ) i 
\label{dsir}
\end{eqnarray}
where $s$, $i$ are the densities of susceptible and infectious individuals, 
respectively, $\gamma $ the recovery rate of the disease and $\beta $ its
transmissibility.  A crucial parameter 
for the behaviour of this model is the so called basic reproductive rate,
$R_0 = \beta / (\gamma +  \mu)$. For $R_0 < 1$ the disease dies out, 
while for $R_0 >1$, the disease is endemic in the population. In this
case, system (\ref{dsir})  has a non trivial endemic 
equilibrium at $s^*= (\gamma + \mu)/\beta $, $i^*= (\mu /\beta)(1/s^* -1)$, 
and the small oscillations around this equilibrium have damping factor $-\mu /(2 s^*)$ and period 
$2\pi/(\sqrt{\mu \beta (1-s^*)-(\mu/2s^*)^2})$.
The Susceptible-Exposed-Infectious-Recovered (SEIR) model considers an
additional class of infected, but not yet infectious, individuals.

In a randomly mixed discrete population of $N$ individuals, SIR dynamics 
may be described as a continuous time Markov process on a population divided
into three classes,  susceptibles,  infectious and recovered.
The state of the system is characterized by the numbers $S$, $I$ and $R=N-S-I$, 
of individuals in each of the three classes, and the events of infection,
death, birth and recovery correspond to the following transitions and 
transition rates
\begin{eqnarray}
T[(S,I) \rightarrow (S-1,I+1)] & = & \beta S I /N\ \ \ \text {infection}  \nonumber \\
T[(S,I) \rightarrow (S-1,I)] & = & \mu S \ \ \ \text {death of a susceptible}  \nonumber \\
T[(S,I) \rightarrow (S,I-1)] & = & \mu I \ \ \ \text {death of an infectious}  \nonumber \\
T[(S,I) \rightarrow (S,I)] & = & \mu (N-S-I)\ \ \ \text {death of a recovered}  \nonumber \\
T[(S,I) \rightarrow (S+1,I)] & = & \mu N\ \ \ \text {birth}  \nonumber \\
T[(S,I) \rightarrow (S,I-1)] & = &  \gamma I \ \ \ \text {recovery},  
\label{trates}
\end{eqnarray}
where $(S,I) \rightarrow (S',I')$ denotes the transition from state $(S,I)$
to state $(S',I')$ and $T[(S,I) \rightarrow (S',I')]$ the corresponding
transition rate.

As detailed in \cite{mckane1}, a good approximation for the power spectrum of 
the stochastic fluctuations around the stationary population numbers can be computed analytically
from the linear Fokker-Planck equation obtained from the next to leading order 
terms in van Kampen's expansion of the master equation associated to (\ref{trates}).
The leading order terms of the expansion yield the deterministic equations (\ref{dsir}),
that describe the behaviour of the system in the limit of infinite populations.

Following this approach, we have computed the power densities 
$P_S$ and $P_I$ of the fluctuations of susceptibles and infectious 
individuals for process (\ref{trates}),  scaled by the square root of 
the system size $\sqrt{N}$, as a function of the angular frequency 
$\omega $, 

\begin{eqnarray}
P_S(\omega) & = &  2 \mu \left (1 - \frac {\gamma + \mu}{\beta } \right )
\frac {(\gamma + \mu)^2 + \omega ^2}{(\omega ^2 - D)^2 + (T \omega)^2} \nonumber \\
P_I(\omega) & = &  2 \mu \left (1 - \frac {\gamma + \mu}{\beta } \right )
\frac {\omega ^2 + \mu^2 (1 - \beta/(\gamma + \mu) + (\beta/(\gamma + \mu))^2)}
{(\omega ^2 - D)^2 + (T \omega)^2} 
\label{PS}
\end{eqnarray}
where $D=\mu (\beta - \gamma - \mu)$ and $T=-\beta \mu /(\gamma + \mu) = -R_0 \mu$,
denoting as before by $R_0$ the basic reproductive rate of the disease.
The parameters $D$ and $T$ are equal to the determinant and to the trace, respectively,
of the linear approximation of (\ref{dsir}) at the endemic equilibrium, and they
have a simple dynamical interpretation: $D$ is the square of the 
frequency of the small oscillations around the equilibrium when the damping is small, 
and $T$ is twice the damping factor of these oscillations.

Equations (\ref{PS}) are independent of $N$ because the dependence of the amplitude
of the fluctuations on system size has been scaled out, and they describe exactly
the stochastic process (\ref{trates}) in the limit of large $N$.
These power spectra are resonant like, showing that the amplitude of the fluctuations
as a function of their time scale is governed by a mechanism of resonance of
internal noise with the system's natural frequency $\sqrt{D}$. The resonance peak 
will be shifted away from $\sqrt{D}$, depending on the values of $T$ 
and on the terms in the numerator of (\ref{PS}),
which are determined by the cross correlation structure of the internal noise.
However, when $T$ is small, the shift in the resonance peak with respect to
$\sqrt{D}$ is also small.

The relevance and universality
of this type of 'endogenous' stochastic resonance for ecological
systems in general was first argued in \cite{mckane2}. The complete analytic description 
of the phenomenon given in \cite{mckane2}, \cite{mckane1} relies on the random
mixing assumption and on the constant recovery rate assumption. The effect
of relaxing these assumptions, crucial in more realistic settings, 
must be tested through stochastic simulations.

\subsection{The SIR model on dynamic small-world networks}

Complex network theory \cite{p41}, \cite{p42}, \cite{p43}  focuses on abstract network 
models such as "small-world" networks, which interpolate between regular lattices 
and random graphs, and scale-free networks, 
exhibiting a power law distribution of the connectivity. In most, if not all, of the 
theoretical work in contact network epidemiology, the host population contact 
network is represented by one of these models. The choice of the right idealized 
network type, which will be disease specific, depends on the availability of 
data on disease causing contacts which is hard to get and difficult to interpret 
\cite{keelingrev}. At this point, an important concern of contact network epidemiology is to 
collect data and build appropriate network models for different transmission 
mechanisms \cite{realnets}.

As a first step to understanding the role of network structure correlations on
the spectrum of stochastic fluctuations, we have modelled the mixing network of
the populations as a small world network \cite{w&s} built over a square lattice
with $12$ nearest neighbours per node.  In these models a fraction of the links of the lattice
is randomized by connecting nodes, with probability $p$, with any other
node. These non-local connections are chosen randomly for each event, instead of
being fixed in a frozen, partially random, link configuration. This version of 
the small world network model, which has been dubbed 'dynamic' or 'annealed'
in the literature (\cite{dynsw}, \cite{szabo}, \cite{meyers}), is motivated by the nature of the occasional
social contacts the model tries to represent.
For $p=0$, each node interacts only with its nearest neighbours on the
lattice, as in ordinary representation of spatial structure.
For $p=1$, the network of interactions is a random graph, where every pair
of nodes, independently of the distance on the lattice between the two nodes,
has the same probability of being connected. Random graphs have the property
that the average path length, or average number of connections  of the 
shortest path between two nodes, is 'small', i. e., of the order of the 
logarithm of the total number of nodes.    
For a range of $p$ between $0$ and $1$ the network exhibits 'small world' behaviour, 
where predominantly local interactions (as in lattices) coexist with a short average  
path length (as in random graphs). Analysis of real networks 
\cite{realnets} reveals the existence of small world 
behaviour in many interaction networks, including  networks of social contacts.

An important consequence of the spatial correlations introduced by
predominantly local contacts is what is called infective
screening, or infective clustering. If all the infected neighbours of an
infected node have many neighbours in common, each of them will 
be connected to a number of susceptibles which is smaller that
the average number of susceptible neighbours per node, and infection
will be less likely than in a randomly mixed population with the same
density of infectious and of susceptibles.

The qualitative analysis of time series of SIR and SEIR stochastic
simulations on dynamic small world networks for different values
of the small world parameter $p$  shows that
indeed the infection process becomes less and less 
efficient as $p$ decreases and infective screening becomes more pronounced 
\cite{cftc1}, \cite{cftc2}.
The global effect of infective clustering may be quantified
in terms of the value of
$\beta_{eff}$, defined as the average number of new
infections per time step and infective individual divided by the density
of susceptibles. For $p=1$, $\beta_{eff}$ coincides with the
transmissibility $\beta$ but, as $p$ decreases, $\beta_{eff}$ also decreases. 
Another qualitative effect of spatial correlations reported in 
\cite{cftc1}, \cite{cftc2} is that they have a major effect on the enhancement of  
the amplitude of stochastic fluctuations, which become more and more 
pronounced as $p$ decreases.

\begin{figure}
\includegraphics[width=\textwidth]{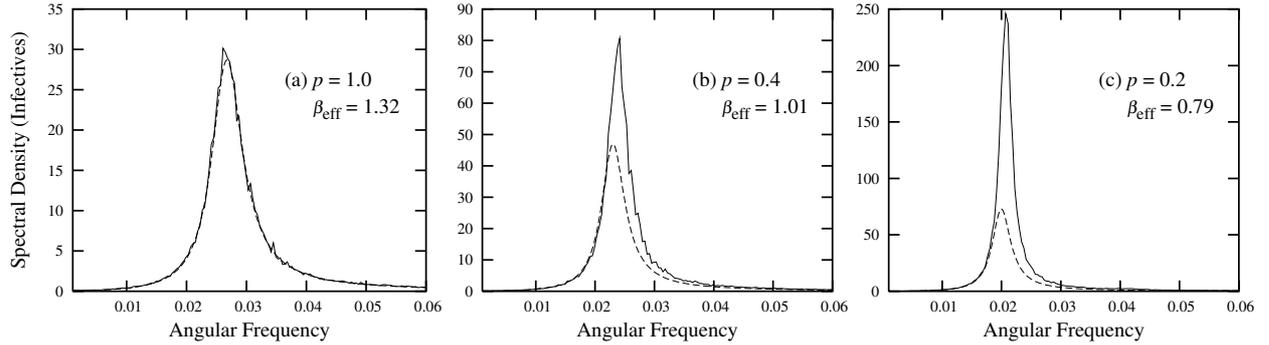}
\caption{Fluctuation power spectra of infectives time series for the SIR model 
on dynamic networks of $N=10^6$ nodes for small world parameter $p$ equal to
(a) 1, (b) 0.4, and (c) 0.2; the networks are built over regular $1000 \times 1000$ 
lattices with 12 first neighbours and periodic boundary conditions.
The full lines are the averaged numerical power spectra of 400 stationary time series, 
each 32768 timesteps long. The dashed lines are the analytic power spectra
given by (3), where for $p$ smaller than 1 the value of the parameter $\beta$ has been 
taken as $\beta_\mathrm{eff}$, defined as the actual number of new infections
per time step and infective individual divided by the density of susceptibles. 
\mbox{Parameter values: $\mu=6 \times 10^{-4}, \; \beta=1.32, \; \gamma=1/8$}.}
\label{fig1}
\end{figure}

In order to quantify the effect of network structure on resonant stochastic
fluctuations, we have computed the power spectrum of SIR time series on dynamic 
small world networks for several values of $p$, and we have compared it with 
the analytic power spectrum given by (\ref{PS}) with $\beta _{eff}$, the effective
transmissibility, instead of $\beta $ (see the Appendix for the details of the
simulations). The results are shown in Figure 1. 
The  full lines are the averaged numerical power spectra of
the stationary time series, and  the dashed lines are the analytic power spectra given by 
(\ref{PS}) with the correction due to the effective transmissibility.
We see that, as $p$ decreases, the resonant
fluctuations are larger and more coeherent. Moreover, in the small world regime,
where local correlations become important, 
this effect is much more pronounced than in the predictions of an effective 
(corrected for infective screening) randomly mixed model, 
represented by the  dashed lines in Figure 1. 
However, this effective model does describe satisfactorily the shift 
to the left in the peak frequency, which means that this can be understood
as a consequence of the reduced effective transmissibility.

The enhanced amplitude and coherence of the fluctuations implies that
a typical time series will exhibit noisy but regular incidence oscillations, as 
shown in Section III.B. These sustained oscillations are of a purely stochastic
nature, and they disappear in the infinite population limit.

The influence of the recovery profile on the behaviour of the system  has been
discussed in \cite{keelingccsize},  \cite{wearing2005}, \cite{lloyd2001b}.
In the standard SIR stochastic model, to which the analytic description
(\ref{PS}) applies, the event of recovery occurs at a fixed
rate during the infection, and the recovery time is exponentially distributed 
around  the average infectious period $\tau = 1/\gamma $, or in other words
uncorrelated.
We have computed the power spectrum of the time series obtained from SIR
simulations in randomly mixed populations where instead of uncorrelated
infection periods
the recovery profile was taken in the opposite limit of constant infection
period $\tau$ or strongly time correlated infections. 
The results are shown in Figure 2. The full line is the averaged numerical 
power spectra of the stationary time series, and  the dashed line is the analytic 
power spectra given by (\ref{PS}) with $\gamma = 1/\tau$. We see that switching to more
realistic (time correlated) recovery profiles leads to a similar effect of enhancing the
amplitude and the coherence of the resonant stochastic fluctuations predicted
by the theory developed in \cite{mckane1}. In addition, there is a large shift to the
right of the dominant frequency of the fluctuations, relative to the
peak frequency predicted for the model with stochastic recovery \cite{mckane1}.
Again, the shift in the peak frequency and its sign can be understood 
in terms of an effective deterministic description, as discussed in Section
III.C. 

\begin{figure}
\includegraphics[width=0.6\textwidth]{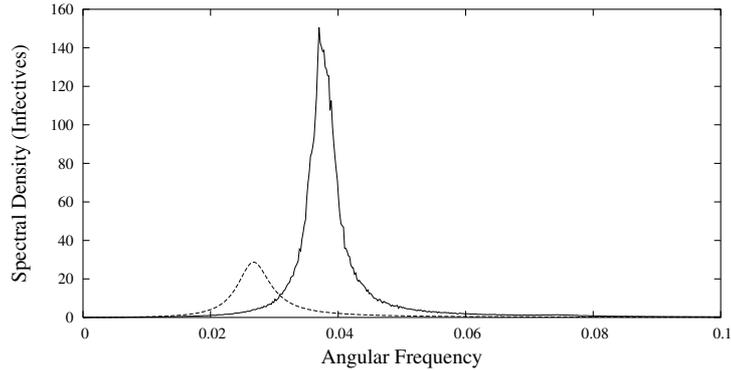}
\caption{Fluctuation power spectra of infectives time series of stochastic SIR 
dynamics with deterministic and stochastic recovery on a randomly mixed population with
$N=10^6$ individuals. Deterministic recovery occurs $\tau$ timesteps after infection. The full line 
is the averaged numerical power spectra of 400 stationary time series,
each 32768 timesteps long. The dashed line is the analytic power spectra given
by (3), 
with $\gamma=1/\tau$.
\mbox{Parameter values: $\mu=6 \times 10^{-4}, \; \beta=1.32, \; \tau=8$}.}
\label{fig2}
\end{figure}

Qualitatively, it is to be expected that correlations in the dynamics
of the system's components should enhance the global density fluctuations. 
To assess quantitatively the effect of spatial and temporal correlations 
on the resonant fluctuations
spectrum in the modelling of infectious childhood diseases, we have performed
systematic simulations in a region of parameter space that includes
the values for measles, chicken pox, rubella, pertussis and mumps according 
to published and estimated data for the pre-vaccination period \cite{b&earn}. The amplitude and 
coherence of the stochastic fluctuations are measured by the overall amplification
$A$, and by the coherence factor $c$, introduced in \cite{mckane1}, where
analytic results for a randomly mixed SIR stochastic model with external infection
and constant recovery rate were presented for the same region of parameter space.
The overall amplification $A$ is the integral over all the frequencies 
of the power density of the scaled
fluctuations, and it is equal to the mean square deviation of the time series 
from the equilibrium values, divided by the system size $N$. The coherence factor $c$ 
is defined as the integral of the power density of the of the scaled
fluctuations on the frequency interval that
corresponds to periods within 10\%  of the peak period $2\pi /\omega_{peak}$, divided by $A$. It
measures the relative contribution to the overall amplification of
stochastic fluctuations that are distributed around the dominant period.

For the randomly mixed case $p=1$ with stochastic recovery, both $A$ and $c$ 
can be computed analytically from equations (\ref{PS}), which are exact 
in the limit of large $N$.  In the remaining cases there
is no analytic description, and the
overall amplification $A$ is 
computed as the ensemble average, over several runs,
of the integral over all sampled frequencies of the power density of the stationary 
time series of the scaled fluctuations. The coherence factor $c$ is computed 
in a similar way on the prescribed frequency interval. The details of the analytical 
and numerical computations are given in the Appendix.

For the model with stochastic recovery, we have also computed, for several values of 
the small world parameter $p$,
the peak shift factor  $s$, defined as the distance
between the actual peak frequency and the natural frequency of the system in 
the deterministic description (\ref{dsir}), $\sqrt{D}$, divided by $\sqrt{D}$.
It measures the relative frequency shift of the dominant frequency due to 
the various ingredients that are missing in the deterministic description 
(resonance with correlated internal noise, and contact network structure).

\section{results}

\subsection{Randomly mixed model: finite size effects}

The randomly mixed model depends on the demographic and epidemiological parameters
$\mu $, $\beta $ and $\gamma $. Following \cite{mckane1}, we have taken $\mu $ 
fixed and equal to $5.5 \times 10^{-5}$, and we have taken the reduced variables 
in the parameter plane $(\mu /\gamma , \beta/\gamma )$. These reduced
parameters have an immediate
epidemiological interpretation, $\mu /\gamma $ measures the acuteness,
or relative time scale, of the disease, and $\beta/\gamma \approx R_0$. 

\begin{figure}
\hspace{-2 cm}
\includegraphics[scale=0.5,viewport=-100 13 820 719,clip]{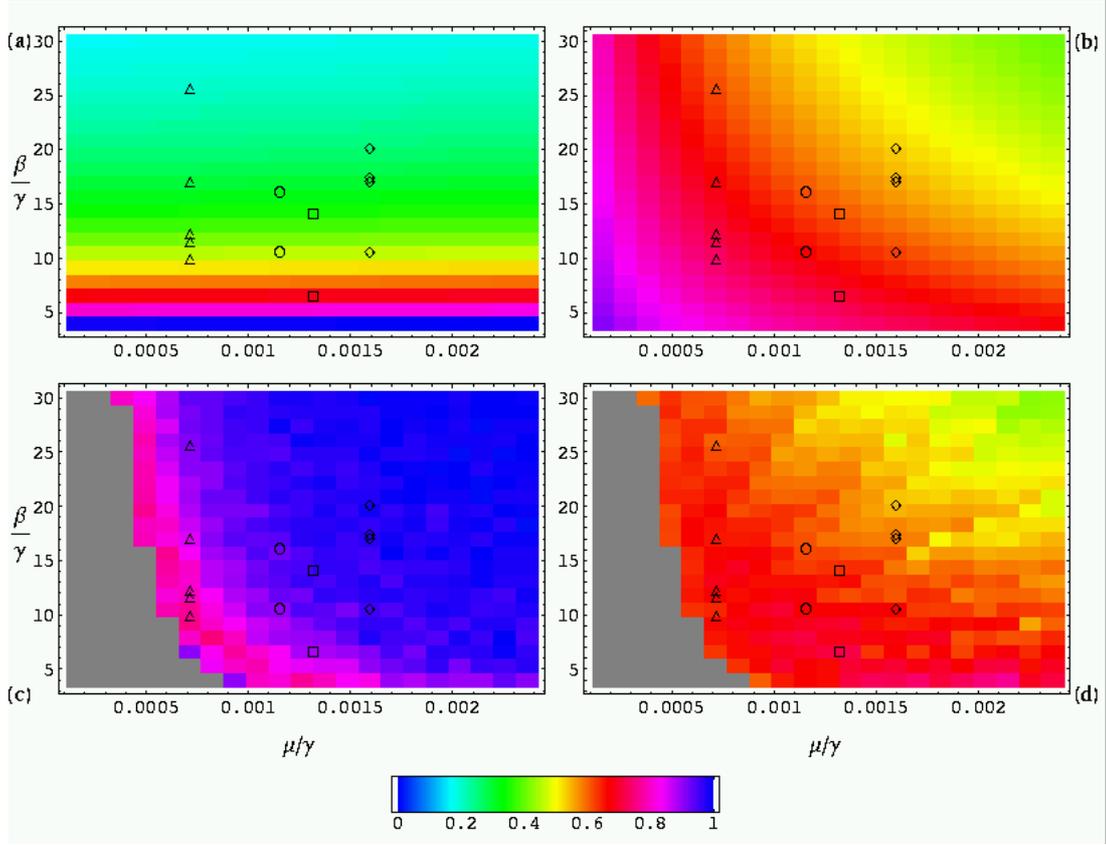}
\caption{Overall amplification and coherence of the randomly mixed model with stochastic recovery 
for $21 \times 21$ points in parameter space. The symbols mark the
parameter values for measles (triangles), chicken pox (circles), rubella (squares) and  
pertussis (diamonds) according to different data sources for the
pre-vaccination period. Analytic results for the amplification $A$
in (a), and for the coherence $c$ in (b).  
The values of (a) are normalized by the largest
overall amplification $0.1902$. In (c) and (d) are plotted the numerical results for the overall 
amplification and coherence obtained from simulations of the stochastic
process (2) on a population of size $N=10^6$. Shown in (c) are 
the values of the analytic plot (a) divided by the corresponding numerical
values. Grey areas denote regions of
parameter space for which less than 200 surviving runs $2^{15}$ timesteps long  were obtained
out of 500 trials.}
\label{figure3}
\end{figure}

The values of the overall amplification
\begin{equation}
A=\frac{1}{\pi}\int_0^{+\infty } P_I(\omega ) d\omega
\end{equation}
and of the coeherence
\begin{equation}
c= \frac{1}{\pi A}\int_{\omega_{peak}/1.1}^{\omega_{peak}/0.9} P_I(\omega ) d\omega
\label{coherence}
\end{equation}
of the infectives power spectrum $P_I$ in (\ref{PS}) are shown in Figure  
3.(a) and 3.(b) for 441 points in parameter space. The values of $A$ are normalized by the 
largest overall amplification $0.1902$. We see that, for the 
basic SIR model, $A$ is essentially determined by $R_0$, 
and that both the overall amplification and the coherence
increase as $R_0$ decreases with $\gamma $ fixed.
For each $R_0 $ the stochastic oscillations 
become more and more coherent as $\gamma $ increases. 
The symbols mark the parameter values for measles, chicken pox, rubella and 
pertussis according to different data sources for the pre-vaccination period 
\cite{b&earn}.
The numerical results for the overall amplification
and coherence obtained from simulations
of the stochastic process (\ref{trates}) on a population of size $N=10^6$ 
are plotted in  Figures 3.(c) and 3.(d) (see the Appendix
for the details of the simulations and numerical computations). 
Each data point in Figure 3.(c) is the ratio of the corresponding analytic  
and numerical values of the overall amplification, while each data point
in Figure 3.(d) corresponds to the numerical value of the coherence as defined
in (\ref{coherence}). The data points shown in grey  correspond to parameter values
where more than 60\% of the 50 000 days long
simulations ended up with zero infectives.
It is clear that for realistic population sizes, in a large region of parameter
space, there are corrections to the analytic results of Figure 3.(a) and 3.(b), and
that the fluctuations for small $\beta $ and large $\gamma $ are larger than 
those predicted 
by the analytic approach.
The numerical power spectra obtained from stochastic simulations for $N=10^6$ and
$N=50 \times 10^6$ for values of $(\mu/\gamma , \beta /\gamma )$ for which
these corrections are more significant are shown in Figure 4.(a). 
The analytic power spectrum given by (\ref{PS}) overlaps the numerical curve
for  $N=50 \times 10^6$ within the resolution of the figure.
A breakdown of the analytic description for
$N=10^6$ can be detected in the overall amplification and in the peak frequency
shift, as well as in the loss of coeherence due to the appearance of a small harmonic
peak. This effect is more pronounced in the presence of spatial correlations,
as shown in Figure 4.(b) and discussed in the following section. 

The limitations of the analytic description (\ref{PS}) are due to the fact that a linear
theory based on van Kampen's method is used to approximate the master equation up to
next to leading order tems, leaving out the contribution of terms of order
$1/\sqrt{N}$ and higher, whose influence on the dynamics becomes more important
as the system size and/or the stability of the stationary state decreases.

\begin{figure}
\includegraphics[width=\textwidth]{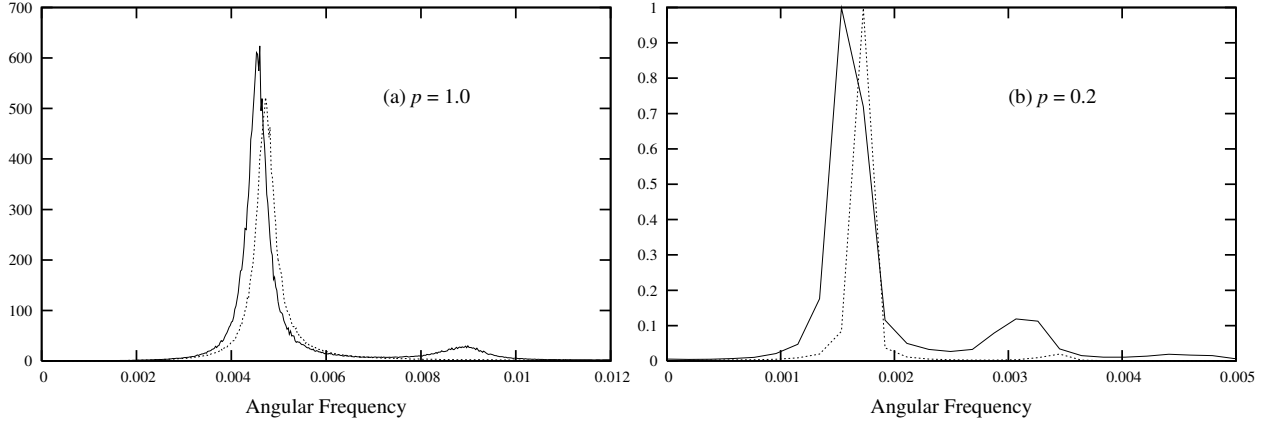}
\caption{(a)  
 Numerical power spectra obtained from stochastic simulations for $N=10^6$ and
$N=50 \times 10^6$ (full and dotted lines, respectively) for values of 
$(\mu/\gamma,\beta/\gamma)=(0.00094,7.9)$. The spectra were taken from 
$2^{18}$, instead of $2^{15}$, timesteps long simulations for increased
resolution in frequency.
The curve for $N=10^6$ shows a larger  
overall amplification and a peak frequency shift, as well as loss of
coherence due to the appearance of a small harmonic peak. (b) Power spectra 
of stochastic simulations on a dynamic small world network with $p=0.2$ and
$N=10^6$ nodes (full line) and $N=50 \times 10^6$ (dotted line), for the parameter
values $(\mu/\gamma,\beta/\gamma)=(0.00182,4.0)$ chosen in the region where
the amplitude of the harmonic peaks is larger. Both plots are scaled to the
largest spectral power ($3960$ for $N=10^6$ and $71000$ for $N=50 \times 10^6$).
The ratio of the heights of the
$2^\mathrm{nd}$ peak to the $1^\mathrm{st}$ peak is approximately $7 \approx \sqrt{50}$
times smaller for the larger system, in line with the expected $1/\sqrt{N}$
dependence.}
\label{fig4}
\end{figure}

\subsection{Small world network: effects of spatial correlations}

To assess the effect of spatial correlations on the resonant fluctuations
spectrum, we have performed, for the same 441 points in parameter space,
systematic simulations of SIR time series on dynamic 
small world networks for several values of $p$ (see the Appendix for the details 
of the simulations). The results for the overall amplification $A$ when $p=0.6$ are shown in Figure 5.a).
The values of  $A$ are normalized by the largest value of the 
overall amplification, which is  $0.2908$. 
With respect to the analytic spectrum (\ref{PS}), the overall amplitude of the fluctuations increases by
a factor of about $1.5$, causing 
more ocasional extinctions. As before, the grey region corresponds to parameter values
where more than 60\% 
of the 50 000 days long simulations ended up with zero infectives.
The dependence of the amplitude of the fluctuations on the epidemiological 
parameters is qualitatively the same
as for the randomly mixed ($p=1$) numerical results of Figure 3.(c). It increases as $R_0 $ decreases,
and, for fixed $R_0 $, it increases as $\gamma $ increases. Close to the 
extinction boundary, especially for low $R_0 $ we find a departure from this general 
trend which can be due a sampling bias, as we have had to select long time
series without disease extinction from a  large ensemble of trial runs. 

\begin{figure}
\hspace{-2 cm}
\includegraphics[scale=0.5,viewport=-100 13 813 367,clip]{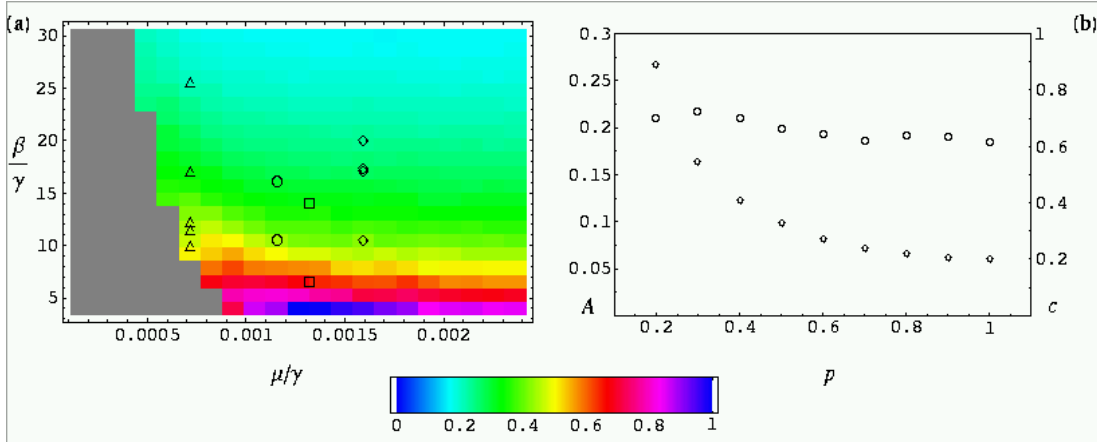}
\caption{(a) Overall amplification of the SIR model with stochastic recovery on a dynamic small world network with
$N=10^6$ individuals and $p=0.6$ (see the Appendix for details of the simulations). The values of the overall
amplification $A$ are normalized in the plot by the largest amplitude ($0.2908$). Again, grey areas denote
regions where only a small percentage of timeseries survive after $2^{15}$ timesteps (each sampled point
requires the survival of at least 200 out of 500 runs). (b) Amplification $A$ (diamonds) and coherence $c$
(circles) for the point $(\mu/\gamma,\beta/\gamma)=(0.00116,17.0)$ as a function of the small world
parameter $p$. The chosen point lies approximately at the center of the region plotted in (a).}
\label{figure5}
\end{figure}

The most relevant effect of spatial correlations is the increase in the amplitude
of the fluctuations as the small world parameter $p$ decreases. The plot of the overall 
amplification $A$ as a function of $p$ is shown in Figure 5.b) (data points plotted with 
diamonds) for parameter values close to those of pertussis (for other disease parameter values 
the behaviour of the amplitude of the fluctuations as a function of $p$ is similar). 
There is a four-fold increase in the 
amplitude of the fluctuations for $p=0.2$ with respect to the randomly mixed case
$p=1$. 

For the same parameter values, the plot of the coherence $c$ as a function of $p$ 
is also shown in Figure 5.b) (data points plotted with circles). There is an increase in the
coherence of the fluctuations as $p$ decreases, which means that as the fluctuations get larger
they also exhibit a more regular temporal pattern.  
For a fixed value of $p$, the coherence is uniformly very high in parameter space
(results not shown), with changes of less than 10\% in a region that includes 
all the parameter values considered for childhood infectious diseases. However, in
a small region close to the $\beta/\gamma =4$ horizontal line and for $p=0.2$ there 
is loss of coherence,  due to the appearance 
of harmonic peaks, see Figure 4.(b), which are more pronounced
than the ones found for $p=1$, and persist for larger values of $N$. 
Although the relative amplitude of the harmonic peaks seems to scale with $1/\sqrt{N}$,
the fact that they are enhanced is due to the presence of spatial 
correlations, and could be related to the existence
of an oscillatory phase for small values of $\mu /\gamma $ in deterministic
SIR models that include, at the simplest level, the effect 
of these correlations \cite{PA}. Indeed, the breakdown of Van Kampen's
$1/\sqrt{N}$ scaling is expected in the oscillatory phase of this model, where the ratio
of the amplitudes of the harmonic peaks is constant, in the infinite size
limit.

The combined effect of the spatial correlations on the amplitude and on the
coherence of the fluctuations can be seen in Figure 6,
where a typical time series for the parameter values of Figure 5.(b) and  
$p=0.2$ is shown. The purely stochastic incidence peaks can be as high as
3500 individuals per day in a population of $N=10^6$, with troughs of one or two hundred
infectious, and a period of recurrence of epidemics can be clearly identified. 
Indeed, the pattern of sustained oscillations shown in Figure 6 is similar to that
found in incidence data for measles in the prevaccination records of english cities
of comparable population size \cite{bolkerweb}, and it is more sharply defined than
that of real time series for most childhood diseases incidence data \cite{olsenschaffer}, 
\cite{ferguson}, \cite{broutin}, \cite{trottier}.

\begin{figure}
\includegraphics[width=0.8\textwidth]{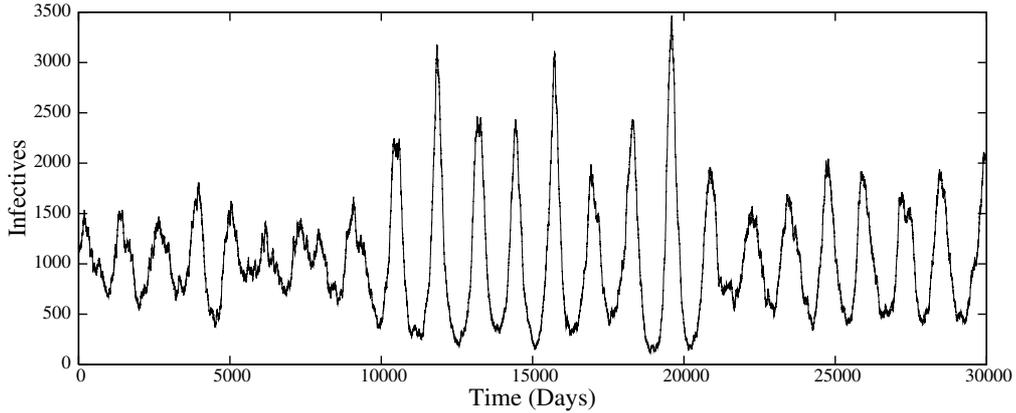}
\caption{Typical incidence time series for a population of $N=10^6$ and small
world parameter $p=0.2$. We have taken
\mbox{$(\mu/\gamma,\beta/\gamma)=(0.00116,17.0)$} as in Figure 5.b).}
\label{fig8}
\end{figure}

\begin{figure}
\includegraphics[width=0.6\textwidth]{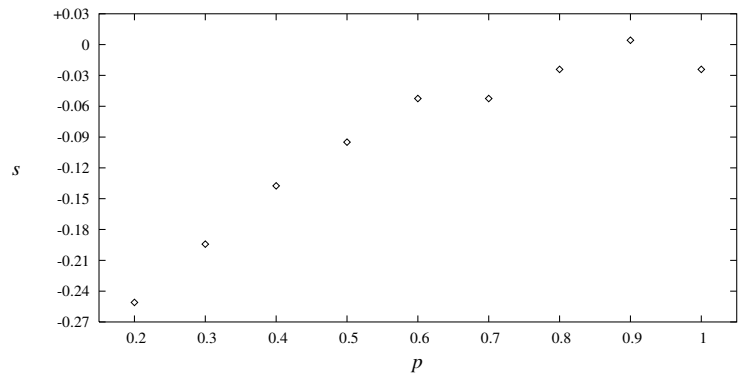}
\caption{The peak shift factor $s$ as a function of the small world parameter $p$. We have chosen
	\mbox{$(\mu/\gamma,\beta/\gamma)=(0.00116,17.0)$}, the same parameter
	values  as in Figure 5.(b).}
\label{fig6}
\end{figure}

Another important effect of contact network structure is the shift in the
dominant frequency of the power spectrum. In Figure 7 we plot, as a function of 
$p$, the peak shift factor $s$, defined as the difference
between the actual peak frequency and the natural frequency of the system in 
the deterministic description (\ref{dsir}), $\sqrt{D}$, divided by $\sqrt{D}$.
Again, we have chosen the same parameter values of Figure 5.b), 
$(\mu/\gamma , \beta /\gamma )=(0.00116,17.0)$, but the results for
other points in parameter space are similar.  We found that, as $p$ decreases, 
the dominant frequency of the 
time series is shifted increasingly to the left. As discussed in Section II, 
this can be undestood as a result of the reduced effective transmissibility
due to clustering of infectives. Since both the overall
amplification and the coherence increase as $p$ decreases, 
the time series of long term simulations will, as shown in Figure 6, exhibit recurrent epidemics
with approximate period close to that given by the dominant frequency of the 
resonant fluctuations. However, this value can be very different from the period that corresponds to
the natural frequency of the system in the deterministic limit
(\ref{dsir}).

\subsection{Recovery profile: effects of time correlations}

\begin{figure}
\hspace{-2 cm}
\includegraphics[scale=0.5,viewport=-100 13 813 367,clip]{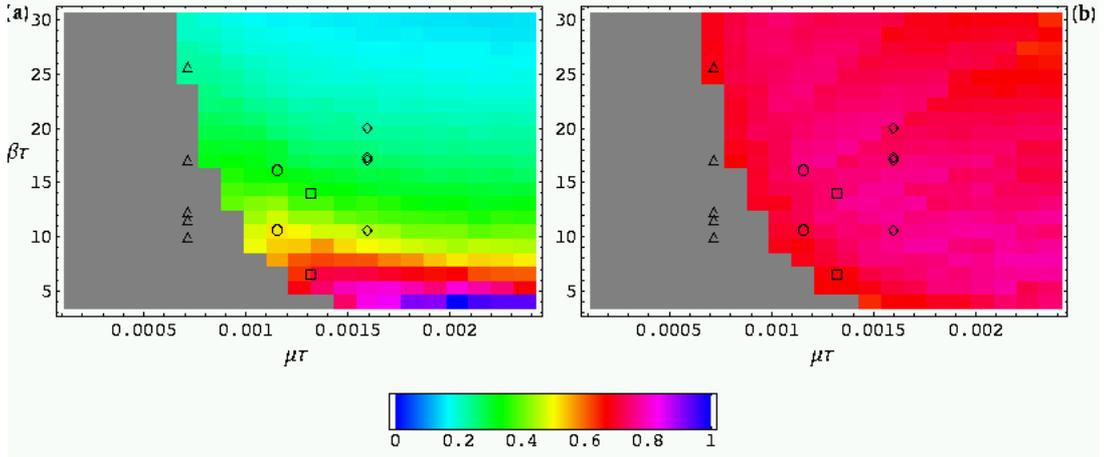}
\caption{Overall amplification (a) and coherence (b) for the randomly mixed SIR model with
deterministic recovery over a population of $N=10^6$ individuals. The values
of (a) are normalized by the largest overall amplification $0.6901$. Again,
grey areas denote regions of
parameter space where less than 200 out of 500 runs survive $2^{15}$ timesteps.}
\end{figure}

The results for the overall amplification and coherence in the 
the randomly mixed SIR model with deterministic recovery are shown
in Figure 8. The values of  $A$ in Figure 8.(a) are normalized by the largest value of the 
overall amplification $0.6901$, which means that the overall amplitude of 
the fluctuations increases by a factor of about $3.5$ with respect to
the model with stochastic recovery. In agreement with the results of Figure 2,
the enhancement of the peak amplitude will in general be even larger,  since, as shown
in  Figure 8.(b), the coherence of the fluctuations for the case of 
deterministic recovery is larger for most parameter values. 
Realistic recovery profiles will therefore be associated with time series
whose fluctuations power spectra will be larger and more sharply peaked
than for constant recovery rate models. The peak frequency shift
relative to this model is towards larger frequencies, and it may
partially cancel the effect of internal noise correlations that shifts the
peak frequency to the left of $\sqrt{D}$, bringing the resulting
dominant frequency closer to  the natural frequency of the system in 
the deterministic description. 

In \cite{lloyd2001b},
the basic SIR model (\ref{dsir}) was modified to include a family,
parametrized by an integer $n$, of infectious periods distributions 
that interpolate between the exponential 
distribution (for $n=0$) and delta distribution associated to deterministic 
recovery (in the limit $n \rightarrow \infty$). Although there is no simple
expression in closed form in this case for the period of the damped oscillations 
close to the system's equilibrium, it was shown that the period decreases as 
$n$ increases.
Our findings are in agreement with the expectation of an increase of the
dominant frequency in the limit of fixed recovery time.

\section{Discussion and Conclusions}

The relevance of the phenomenon of resonance with internal noise for the
understanding of the long term dynamics of childhood infections, with or
without the presence of seasonal forcing, has been uncovered in \cite{mckane1}.
The power spectrum of the fluctuations of the stochastic process of 
infection in a population has been given a complete analytic description, 
under the following basic assumptions:\par

1. There is external infection at a small constant rate
\par 
2. The population is randomly mixed and internal infection follows a law of
mass-action with constant transmissibilty
\par 
3. Recovery from the disease occurs at a constant rate
\par 
4. The behavior of the fluctuations is well described by the 
lowest order van Kampen expansion of the master equation, 
\par \noindent
that correspond to a stochastic open SIR model and large population
sizes. 

We have extended the analytic results of \cite{mckane1}
to a closed stochastic SIR model, with no external infection, 
that models an isolated population, and compared these with the results of
extensive numerical simulations. Due to stochastic extinctions,
simulations of the long term behaviour of this model require large population
sizes, and we took $N=10^6$ as a typical value for the population size.
We have found a good agreement between the analytic description and the
results of the simulations for most values of the parameters. However,
in a region of parameter space that includes in particular measles,
we have found that there are significant finite size corrections to the analytic power
spectrum of the fluctuations, which means that assumption 4
is fullfilled only for much larger population sizes (indeed for 
population sizes of $N=5 \times 10^7$ we have
found good agreement between the analytic description and the
results of the simulations, over the whole range of parameters).   

We have then investigated the effects on the fluctuation power spectrum of relaxing
2 or 3 in order to account for more realistic contact networks of the population
or disease recovery profiles. These are examples 
spatial and temporal correlations that are always present in real
interacting systems. 
In either case, the approximate analytic description is no 
longer valid, and we must resort to systematic simulations.
However, we have found fluctuation power spectra that are 
dominantly resonant like, which suggests that the basic mechanism at play is
still resonance with demographic stochasticity. 

Instead of assuming 2, we have modelled the mixing network of the 
populations as a dynamic small world network, and considered several different
values of the small world parameter $p$ that measures the degree of randomness of
the contact network. We have found that, as $p$ decreases, the resonant fluctuations
become larger and more coherent over the whole range of parameters that are relevant for
the description of childhood infections, and the dominant frequency of these
fluctuations is shifted towards smaller frequencies.  

Instead of 3, we have considered the opposite (strongly correlated) limit of deterministic 
recovery at the end of the infectious period. Clearly, realistic recovery profiles will lie
between these two extremes. We have found, again,  resonant fluctuations that 
are much larger, and more coherent than when recovery occurs randomly at a constant rate,
and that the dominant frequency is shifted to larger frequencies.

The main conclusion is then that the importance of the phenomenon reported in 
\cite{mckane1} to the description of the long term dynamics of childhood diseases
is enhanced when the model is modified to include more realistic assumptions
(correlations), either on
the populations contact patterns (spatial correlations) or on the disease recovery 
profile (time correlations).
Our results apply to a large region in the epidemiological and demographic
parameter space, and in this sense they are applicable in general to the 
modelling of infectious disease dynamics. They show that stochasticity may
play the leading role in determining the incidence temporal patterns,
through resonance of internal noise with the dynamics that governs the system
in the limit of an infinite population. 

Our results also  show that the analytic theory
developed in \cite{mckane1} provides an overall good description for well mixed
populations and diseases with gradually decaying recovery profiles, and 
a useful guideline for the case of populations with non trivial contact networks.
When these contact networks include a large fraction, of 50\% or more, of random 
connections, an effective theory based on the analytic treatment of the randomly mixed
case with a correction of the transmissibility to account for infective screening
gives a good description of the incidence fluctuations spectrum. In highly correlated
contact networks, however, the overall amplitude of the fluctuations is much larger than
that predicted by this effective theory.

The qualitative picture for the dependence of demographic
stochasticity on the system parameters that emerges from our results
is more subtle than what one would expect from linear perturbation
analysis, either of the deterministic model (\ref{dsir}) or even of the
full stochastic description (\ref{trates}). 
First, in the presence of correlations, the dominant frequency
of these resonant organized fluctuations differs significantly
 from the natural frequency of the deterministic description. 
The fact that this naive expectation may seem to lead to a good  description,
as argued in \cite{b&earn}, must be attributed to the cancelation of several
missing effects. Second, the amplitude of these fluctuations decreases with $N$, 
or more precisely with $\sqrt{N}$, but finite size effects will be important
for realistic population sizes, of the order of $10^6$, except for diseases with 
an epidemiological profile similar to that of pertussis. Third, the amplitude of 
the fluctuations increases when $R_0 $ decreases, and, for fixed $R_0$, it decreases
as $\gamma $ decreases. This additional 
dependence of the amplitude of the fluctuations on the average infectious period 
is unaccounted for by the analytic description (\ref{PS}).
 
On more conceptual grounds, the finding that in finite, discrete populations internal noise together
with correlations produces sustained incidence oscillations of significant
amplitude all over the parameter region that includes childhood infectious
diseases is of importance for the long-standing controversy in 
epidemiology and ecology as to the driving mechanisms of the pervasive
noisy oscillations observed in these systems \cite{grenfellborn}. Whether
these are mainly intrinsic or external seems to depend not only on the
model's nonlinearities but also on the correlations between the systems's
units, which most traditional approaches neglect.

As the need for spatial models of infectious disease transmission is 
increasingly acknowledged, there are different modelling strategies
that try to reconcile explicit spatial representation with computational
costs and the informatin available on the patterns of contact of the
populations (see \cite{riley} and refences therein). In the context of 
this discussion, it should be stressed that the kind of intrinsic 
stochastic effects highlighted in the present paper is specific of models
that explicitely represent individuals (or small units such as households), 
and that computationally lighter, coarse-grained models will miss this phenomenology.

\section{Appendix}

All simulations were carried out over a population of individuals arranged on a square lattice
with periodic boundaries where each site connects to its 12 closest neighbours. A random fraction
$p$ of these local links are replaced at each step with random long-distance links. The
population size was $10^6$ individuals for all simulations, with the exception of the results
of Figure 4 where a population of $50 \times 10^6$ individuals was also considered.

The simulations implemented the stochastic process of the SIR model, or its
modification to include deterministic, instead of stochastic, recovery, on this network.
For the case of stochastic recovery, we have used the efficient algorithm 
for stochastic processes in spatially structured systems
described in~\cite{BKL}. Local and long range infections are dealt with separately,
and a local (resp. long range) infection event occurs with probability $1-p$
(resp. $p$). For $p=1$, the presence of spatial structure is irrelevant, and
the algorithm reduces to the application of the method of Gillespie~\cite{Gillespie}
to the stochastic process (\ref{trates}). When $p<1$ and local infections may occur,
their probability depends on the number of infected neighbours of each susceptible
node, and so susceptible nodes with $k$ infected neighbours, $k=0, ..., 12$ are
treated as separate classes.
For the case of stochastic recovery, this algorithm is extended
with the inclusion of a linked list of infected nodes.
Each entry in the list records the position of an infected node in the lattice, 
along with the timestep number (counting from 0 at the start of the run) at which this
particular node should recover (in case it has not been removed previously by a stochastic death transition).
The list is ordered by recovery timestep number; that is, it always contains the oldest nodes at the start.
In this way, recovery is efficiently performed by checking, at the beginning of each timestep, which nodes
are due for recovery and replacing them in the lattice with recovered nodes

The numerical results of Figures 3, 5 and 8
were obtained through systematic simulations of the SIR model for a
set of $21 \times 21$ evenly spaced points covering the $(\mu/\gamma,\beta/\gamma)$ plane, for
the region depicted in the figures (and keeping $\mu=5.5 \times 10^{-5}$ fixed). Four separate
sets of simulations were carried out, one for each of the specific cases considered in the main text, namely,
the SIR model with stochastic recovery for $p=1.0$ and for $p=0.6$, 
the SIR model with stochastic recovery for $(\mu/\gamma,\beta/\gamma)=(0.00116,17.0)$
and several values of $p$, and  the SIR model with deterministic recovery and $p=1.0$.

For each sample point, 500 independent runs of the model were taken, each lasting for 50000
timesteps or days, of which only the last $2^{15}$ are used; approximately
17000 timesteps are discarded at the begining, to allow each run to `settle down' to
its steady state. If less than 200 of these 500 runs survive (that is, finish with more than zero
infectives), then this particular sample point is discarded (shown as grey in the plots); otherwise,
the power spectral densities of the surviving runs are computed (using an FFT routine) and averaged
together. The remaining power spectrum plots shown in the Figures were obtained in
the same way, except those of Figure 4.(a), in which
$2^{18}$ timesteps long stationary timeseries were used, 8 times longer
than those used for other plots, to provide the increased frequency resolution necessary
to check the perfect agreement with the the analytical prediction (3) for $50 \times 10^6$
individuals.

From the final averaged spectrum, the quantities $A$, $c$ and $s$ are computed. The overall
amplification, which is defined as $A={1 \over \pi} \int_{0}^{+\infty} P_I(\omega) \, \mathrm{d}
\omega$, where $P_I(\omega)$ is the power spectrum density of the scaled fluctuations,
is numerically approximated by a sum over the squared moduli of the scaled coefficients $Z_k/\sqrt{N}$
of the discrete Fourier transform of the timeseries. Since the sampling interval is $1$ day, 
the band width of the signal is $\omega \in [-\pi, \pi]$ , and we have

\begin{displaymath}
A \approx {1 \over \pi} \int_{0}^{+\pi} P(\omega) \, \mathrm{d} \omega={1 \over L}
	\int_{0}^{L} P(\pi k/L) \, \mathrm{d} k \approx {1 \over L} \sum_{k=0}^{L-1} {|Z_k|^2 \over N}
\end{displaymath}
where $k$ is the integer index of the Fourier coefficient and $L$ the sample length (here $L=2^{15}$). 
The coherence is defineded as $c=A_p/A$, where $A_p$ is the integral of the power spectrum density 
of the scaled fluctuations over the frequency range that corresponds to $\pm 10 \%$ of the 
period of the dominant peak. The numerical value of $A_p$ is obtained in the same manner as $A$
from the squared moduli of the scaled coefficients $Z_k/\sqrt{N}$, replacing the summation limits 
appropriately. 

The position of the dominant peak $\omega_{peak}$ is found from the largest Fourier coefficient, 
and is also used to compute the  shift factor $s = (\omega_{peak} - \sqrt{D})/\sqrt{D}$ shown in
Figure 7.

\section{Acknowledgements}

Financial support from the Foundation of the University of Lisbon 
and the Portuguese Foundation for Science and 
Technology (FCT) under contracts POCI/FIS/55592/2004
and POCTI/ISFL/2/618 is gratefully 
acknowledged.

\end{document}